\documentclass[reprint,a4paper, amsfonts, amssymb, amsmath, reprint, showkeys, nofootinbib, twoside, superscriptaddress]{revtex4-2}

\usepackage[english]{babel}
\usepackage[utf8]{inputenc}
\usepackage[colorinlistoftodos, color=green!40, prependcaption]{todonotes}
\usepackage{amsthm}
\usepackage{mathtools}
\usepackage{physics}
\usepackage{xcolor}
\usepackage{graphicx}
\usepackage[left=23mm,right=13mm,top=35mm,columnsep=15pt]{geometry} 
\usepackage{adjustbox}
\usepackage{placeins}
\usepackage[T1]{fontenc}
\usepackage{lipsum}
\usepackage{csquotes}

\input{Commandes.sty}

\usepackage{aas_macros}

\usepackage{amsmath}
\usepackage{amssymb}
\usepackage{amsfonts}
\usepackage{mathtools}
\usepackage{booktabs}
\usepackage{stmaryrd}
\usepackage{mathrsfs}
\usepackage{multirow}
\usepackage{centernot}

\usepackage{fancyvrb}
\usepackage{graphicx}
\usepackage{subfig}
\usepackage{wrapfig}
\usepackage{caption}
\usepackage{cases}
\usepackage{tikz}
\usepackage{eurosym} \def\{\euro{}} 
\usepackage{cmlgc}
\usepackage{color}
\usepackage{empheq}
\usepackage[most]{tcolorbox}
\usepackage{booktabs}
\usepackage{textcomp} 
\usepackage{soul}
\usepackage{natbib}
\usepackage{graphicx}
\usepackage{txfonts}

\usepackage{url}
\definecolor{linkcolor}{rgb}{0,0,0}
\definecolor{linkcolorurl}{rgb}{0,0,1}
\usepackage[pdftex,colorlinks=true,
pdfstartview=FitV,
linkcolor= blue,
citecolor= blue,
urlcolor= blue,
hyperindex=true,
hyperfigures=true]
{hyperref}
\hypersetup{linktoc=page}

\usepackage{siunitx}
\DeclareSIUnit\erg{erg}
\DeclareSIUnit\gauss{G}
\DeclareSIUnit\arcsec{as}
\DeclareSIUnit\parsec{pc}
\DeclareSIUnit\rad{rad}
\DeclareSIUnit\jansky{Jy}

\begin{document}
\title{Images of magnetospheric reconnection-powered radiation around supermassive black holes}

\defcitealias{Crinquand_2021}{C21}

\author{Benjamin Crinquand}
    \email[Correspondence email address: ]{bcrinquand@princeton.edu}
    \affiliation{Univ. Grenoble Alpes, CNRS, IPAG, 38000 Grenoble, France\\}
    \affiliation{Department of Astrophysical Sciences, Peyton Hall, Princeton University, Princeton, NJ 08544, USA}
\author{Benoît Cerutti}
    \affiliation{Univ. Grenoble Alpes, CNRS, IPAG, 38000 Grenoble, France\\}
\author{Guillaume Dubus}
    \affiliation{Univ. Grenoble Alpes, CNRS, IPAG, 38000 Grenoble, France\\}
\author{Kyle Parfrey}
    \affiliation{School of Mathematics, Trinity College Dublin, Dublin 2, Ireland}
\author{Alexander Philippov}
    \affiliation{Center for Computational Astrophysics, Flatiron Institute, 162 Fifth Avenue, New York, NY 10010, USA}
    \affiliation{Department of Physics, University of Maryland, College Park, MD 20742, USA}

\date{\today} 

\begin{abstract}

Accreting supermassive black holes can now be observed at the event-horizon scale at millimeter wavelengths. Current predictions for the image rely on hypotheses (fluid modeling, thermal electrons) which might not always hold in the vicinity of the black hole, so that a full kinetic treatment is in order. In this letter, we describe the first 3D global general-relativistic particle-in-cell simulation of a black-hole magnetosphere. The system displays a persistent equatorial current sheet. Synthetic radio images are computed by ray-tracing synchrotron emission from nonthermal particles accelerated in this current sheet by magnetic reconnection. We identify several time-dependent features of the image at moderate viewing angles: a variable radius of the ring, and hot spots moving along it. In this regime, our model predicts that most of the flux of the image lies inside the critical curve. These results could help understand future observations of black-hole magnetospheres at improved temporal and spatial resolution. 

\end{abstract}

\maketitle

\paragraph{Introduction}

At least part of the formidable energy released by active galactic nuclei (AGN) may originate from the rotational energy of the central black hole, extracted by the means of strong magnetic fields threading its event horizon~\citep{Blandford_1977}. These magnetic fields are sustained by currents flowing in an accretion flow, which in the case of a low-luminosity AGN is hot, geometrically thick, and collisionless. Recently, for the first time, the Event Horizon Telescope (EHT) collaboration was able to spatially resolve the immediate vicinity of the low-luminosity AGN M87*. The image shows a circular asymmetric ring encompassing a depression in brightness~\citep{EHT_1}. From the high polarization fraction and brightness temperature of the radiation, we know that the light captured by the EHT is optically thin synchrotron emission emitted by mildly relativistic leptons. These measurements are consistent with the scenario of dynamically important poloidal magnetic fields at event horizon scales, thus favoring the ``magnetically arrested disk'' (MAD) scenario for the accretion flow~\citep{EHT_2021a,EHT_2021b}.

However, the location of the main emission sites, as well as the mechanisms causing plasma heating and particle acceleration,  are still poorly constrained. A widely adopted approach has been to interpret the emission as synchrotron radiation by thermal~\citep[e.g.][]{Moscibrodzka_2016,EHT_5} or non-thermal~\citep{Davelaar_2019,Scepi_2021,Fromm_2022} electrons, energized for example by subgrid kinetic turbulent dissipation or magnetic reconnection. General-relativistic magnetohydrodynamic (GRMHD) simulations can be educated with prescriptions for electron heating used to model the local synchrotron emissivity, which allows one to synthesize images when coupled with GR ray-tracing algorithms. In this framework, the image should prominently display a black-hole shadow determined by the properties of the spacetime~\citep{Falcke_2000,Narayan_2019, Bronzwaer_2021}. It is still unclear, though, to what extent the features of the image are related to the astrophysical details of the emission~\citep{Gralla_2019,Nalewajko_2020,Vincent_2021,Lockhart_2022}.

Recently, several GRMHD studies of MAD disks (\citet{Ripperda_2020,Chashkina_2021} in 2D and \citet{Ripperda_2022} in 3D) have reported dynamical transitions from an accreting state, during which magnetic flux is accumulated onto the horizon, to a highly magnetized quasi-force-free state, which is characterized by a stark drop in the accretion rate and the existence of a thin equatorial current sheet. This configuration favors fast dissipation of the magnetic energy accumulated in the magnetosphere through magnetic reconnection in the current sheet, triggering efficient particle acceleration and high-energy nonthermal radiation~\citep{Sironi_2014,Guo_2014}. Thus, it has been suggested that gamma-ray flares from radio galaxies (such as M87*) could naturally occur when the central supermassive black hole is in this quasi-force-free state, hereafter labeled ``high-energy flaring state''.

In this regime, with the innermost zone almost depleted of electron-ion plasma, predictions for the image based on GRMHD do not apply, whereas kinetic simulations can accurately capture the physics and dynamics of the highly magnetized and collisionless plasma. Global kinetic particle-in-cell (PIC) simulations of black-hole magnetospheres have only recently begun to be carried out~\citep{Parfrey_2019, Crinquand_2020, Crinquand_2021, Bransgrove_2021, El_Mellah_2021}. \citet{Crinquand_2021} (hereafter, \citetalias{Crinquand_2021}) previously studied the dynamics and high-energy gamma emission of the equatorial current sheet that develops generically within the ergosphere from a large class of ordered magnetic configurations~\citep{Komissarov_2004}. Conversely, in this letter, we characterize the low-energy emission at radio wavelengths from a black hole in a high-energy flaring state, embedded in a quasi-force-free magnetosphere, and discard any contribution from the accretion flow.

 
\paragraph{Simulations}

We use the GRPIC code \texttt{GRZeltron} to simulate magnetospheric pair plasma. The background spacetime is described by the Kerr metric with a dimensionless spin parameter $a=0.99$. We use spherical horizon-penetrating Kerr-Schild coordinates $\left( r, \theta, \varphi \right)$, and model self-consistent plasma injection by including inverse Compton scattering and photon-photon annihilation~\citep{Crinquand_2020}. We have run both a 2D and a 3D simulation, using the setup described in~\citetalias{Crinquand_2021} with initially paraboloidal magnetic field lines switching polarity at the equator. The 3D simulation has a resolution of $1024 \left( r \right) \times 256 \left( \theta \right) \times 512 \left( \varphi \right)$, whereas the 2D one has $1536 \left( r \right) \times 1024 \left( \theta \right)$. The simulation domain is $r \in [0.9 \, \rh, 10 \, \rg], \theta \in [0.1,\pi-0.1], \varphi \in [0, 2 \pi]$, with $\rg$ the gravitational radius of the black hole and $\rh$ the radius of the event horizon. Our fiducial 2D and 3D runs do not have synchrotron cooling. Because of a higher numerical cost, the values of the dimensionless input parameters are slightly less realistic in 3D than in 2D: the fiducial Larmor radius $r_\mathrm{L}$ of particles is $r_\mathrm{L} = 10^{-4} \, \rg$ in 3D instead of $2 \times 10^{-6} \, \rg$ in 2D, whereas the background soft photon energy is $\varepsilon _0 = 10^{-2} \, \me c^2$ instead of $10^{-3}$. In both cases, the scale separation is necessarily smaller than that of a realistic AGN. M87*, for example, is characterized by a ratio $r_\mathrm{L} / \rg \approx 10^{-14}$, which is currently unfeasible numerically. We have also confirmed that 2D simulations are not sensitive to the precise initial distribution of plasma and photons.



\begin{figure}[t!]
    \centering
    \resizebox{1.0\hsize}{!}{\includegraphics{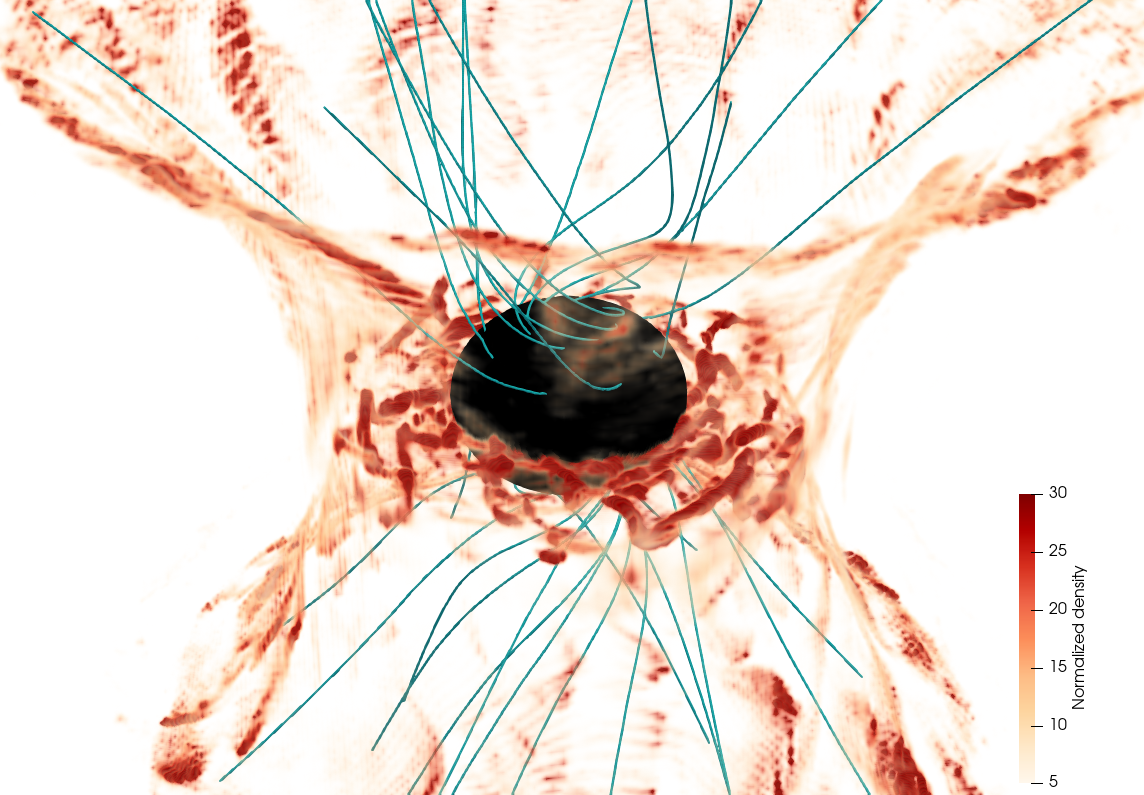}}
	\caption{Snapshot of the total plasma density, normalized by the fiducial Goldreich-Julian density, at $t=40 \, \rg/c$. Magnetic field lines are represented as blue solid lines. The black surface marks the event horizon.}
	\label{fig:3D}
\end{figure}

The general behavior of the 3D simulation is very similar to that of the 2D one. The magnetic field develops a strong toroidal component, with opposite signs above and below the equator. Thus, a current sheet is needed to support the associated discontinuity. Pairs are accelerated by nonideal electric fields in the equatorial plane, triggering high-energy photon upscattering and even stronger pair creation. The densities are significantly larger in the current sheet than in the polar regions. The initial current sheet quickly fragments into flux ropes (see Fig.~\ref{fig:3D}). The 3D evolution of the reconnecting current sheet is governed by the tearing instability, which produces flux ropes elongated in the direction orthogonal to the reconnecting field, and the drift-kink instability, which can corrugate the sheet along the reconnecting field~\citep{Sironi_2014,Cerutti_2014}. Besides, the flux ropes are orbiting around the black hole and are constantly being sheared by differential rotation, and plasma can be pushed away from the reconnecting sheet and ejected along the separatrix (the last open field line connected to the black hole). These competing effects drive the current sheet toward a state of twisted and disordered flux ropes. It should be pointed out that our simulation ran for around $40 \, \rg/c$, which is too short to observe the full cyclic behavior observed in~\citetalias{Crinquand_2021}, during which a giant magnetic island, formed between the current sheet and the separatrix, replenishes the black hole with magnetic flux. Although these flux ropes could lose coherence for longer integration times, we note that the simulation has run in steady state for about $20 \, \rg /c$, which is longer than the characteristic growth time of the tearing and drift-kink instabilities which could disrupt it~\citep{Zenitani_2007} and longer than the orbital period at the event horizon  ($4 \pi / \omega_\mathrm{h} \sim 14 \, \rg /c$, with $\omega_\mathrm{h} = a c / 2 \rh$).

\paragraph{Ray-tracing}   

\begin{figure*}[ht!]
    \centering
    \resizebox{\hsize}{!}{\includegraphics{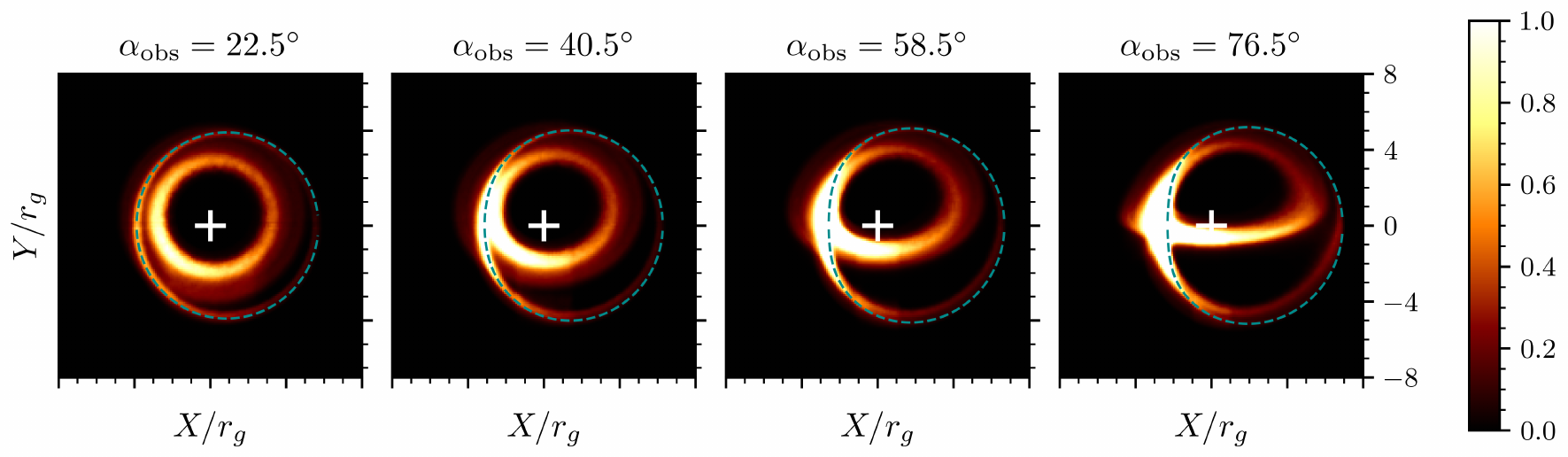}}
	\caption{Time-averaged synchrotron images from the 2D simulation. The flux is in arbitrary units. The three columns show three different viewing angles $\alpha_\mathrm{obs}$, measured with respect to the spin axis of the black hole. The white cross marks the position of the black hole. The blue dashed line denotes the critical curve for $a=0.99$ at each viewing angle.}
    \label{fig:mean}
\end{figure*}


We use the public ray-tracing code \texttt{geokerr}~\citep{Dexter_2009} to perform geodesic integration of synchrotron photons from the emission point to a distant observer's screen, in post-processing (see the Supplemental Material, which includes Ref. \citep{Carter_1968}). When analyzing the 2D simulation, we focus on the time dependence on the image. By virtue of axisymmetry, deposition of photons on the observer's screen is independent of the observer's azimuthal viewing angle. By contrast, when processing the 3D simulation, we effectively take a snapshot of the simulation but allow for different azimuthal viewing angles $\varphi_\mathrm{obs}$. This way, we can separately assess the effects of time variability (inaccessible to the short 3D simulation) and nonaxisymmetry of the current sheet (inaccessible to the 2D axisymmetric simulation) on the image. 

We only show the contribution of the current sheet. Indeed, synchrotron emission by particles accelerated in the polar gaps is unlikely to be observable. Only a small fraction of the Blandford-Znajek electromagnetic power is dissipated in these gaps, which probably achieve plasma densities too low to yield significant emission. Besides, unlike particles accelerated in the current sheet, particles emitted in the polar regions have low pitch angles, implying a comparatively low synchrotron emission. We have also run a 2D simulation with synchrotron cooling turned on, where the synchrotron radiation-reaction force is rescaled, so that the characteristic cooling time of an electron should be a few simulation time steps~\citep{Cerutti_2016}. In that case, the morphology of the image is practically unchanged whereas the polar cap emission is quenched. This is consistent with previous PIC studies of magnetic reconnection with synchrotron cooling~\citep{Cerutti_2013,Kagan_2016,Hakobyan_2019}: particles are accelerated near X points where the magnetic field vanishes, allowing particles to reach high energies before escaping the reconnecting layer and cooling.

\paragraph{Averaged images}


Fig.~\ref{fig:mean} shows images from our 2D simulation averaged over $100 \, \rg/c$, which are very similar to the images from our 3D simulation averaged over the azimuthal viewing angle. The leftmost panel represents a view close to face-on, which is the most relevant to the M87* system, assuming that the spin of the black hole is aligned with the large-scale jet~\citep{Walker_2018}. In general, the image is made up of two distinct components: an inner and an outer ring. The rings' brightness is higher on the left side of the image, as a result of relativistic beaming: this is mainly a geometrical effect due to the Kerr metric, rather than Doppler boosting. The inner ring is the direct image of the current sheet, whereas the outer ring is the lensed image. On each image, we have plotted the critical curve corresponding to each inclination, i.e. the set of points on the screen hit by photons which have orbited an arbitrarily large number of times around the black hole close to bound, marginally stable spherical orbits~\citep{Cunningham_1973,Teo_2003}. This curve matches well the outer ring emission, confirming its interpretation as the lensed image of the current sheet. However, the lensed image accounts for only a small portion (around $10 \%$) of the flux of the image. Our model hence predicts that in a high-energy flaring state, most of the flux of the image lies inside the critical curve. In this configuration, the black-hole shadow prediction breaks down because most of the emission comes from within the ergosphere and is not spherically distributed. We find that a major fraction of the current sheet radiation originates from a very compact zone, within $2-3 \, \rg$. Both rings become thinner as the observing frequency increases, since the emission originates from the most energetic particles which have been accelerated deeper within the current sheet. 


\begin{figure*}[ht!]
    \centering
    \resizebox{\hsize}{!}{\includegraphics{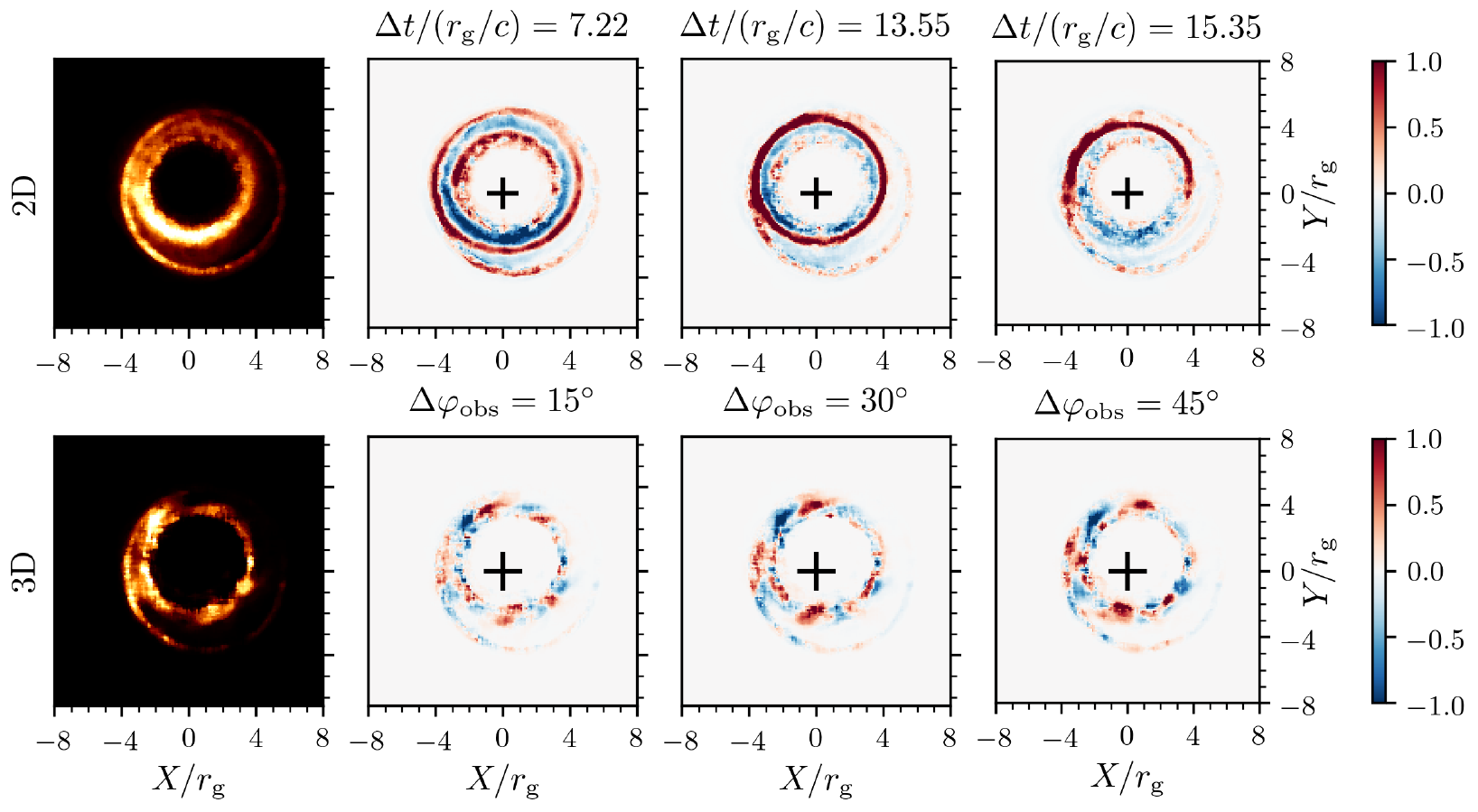}}
	\caption{Top row: Snapshot of the image from the 2D simulation (leftmost panel) and differential maps illustrating the time evolution. The three maps show the difference between successive snapshots at later times and the first one. Bottom row: Image from the 3D simulation at a given azimuthal viewing angle $\varphi_\mathrm{obs}$ (leftmost panel) and differential maps illustrating the dependence on $\varphi_\mathrm{obs}$. The three maps show the difference between images at higher $\varphi_\mathrm{obs}$ and the first one.}
    \label{fig:variability}
\end{figure*}


\paragraph{Variability} 

The first row of Fig.~\ref{fig:variability} illustrates the time variability in the 2D axisymmetric simulation, for a viewing angle $\alpha_\mathrm{obs}=22.5^\circ$. Images are labeled using the observer's Boyer-Lindquist coordinate time. At a certain time during the simulation, a bright circular feature appears between the inner and outer rings (top left panel). As time elapses, this feature shrinks in size until it overlaps the inner ring. Its typical lifetime is about $10 \, \rg /c$. By computing the synchrotron emissivity of the plasma, we can relate this emission to a giant magnetic island accretion event (see ~\citetalias{Crinquand_2021}). The second row of Fig.~\ref{fig:variability} illustrates the dependence on $\varphi_\mathrm{obs}$ of images from the 3D simulation, for the same viewing angle. Both the inner and outer rings show nonaxisymmetric features in the form of brighter spots. They are more pronounced on the inner ring, which is less impacted by gravitational lensing. The presence of hot spots remains visible even after blurring the image at a resolution $\simeq 5 \, \rg$. These hot spots can be traced back to the different equatorial orbiting flux ropes. They radiate relatively isotropically, so that every flux rope is visible regardless of $\varphi_\mathrm{obs}$. This is why the positions of the hot spots along the ring shift continuously clockwise with increasing $\varphi_\mathrm{obs}$ in Fig.~\ref{fig:variability}. The hot spots appear to move counterclockwise along the ring, with a period $\sim 4 \pi / \omega_\mathrm{h} \sim 14 \, \rg/c$, to a fixed observer at a given $\varphi_\mathrm{obs}$.

Although the ratio between $\rg$ and the electron plasma skin depth is unrealistically small in these simulations, the hierarchical merger of magnetic islands should proceed until plasmoids with macroscopic sizes are formed. The final number of synchrotron-emitting flux ropes, and hence of hot spots on the image, is determined by their characteristic escape time from the merging region~\citep{Cerutti_2021}. The flux ropes are constantly reformed within the current sheet to compensate for their escape, along the separatrix or into the black hole. \citet{Cerutti_2021} predict the final number of macroscopic plasmoids to be $\sim \pi / \beta_\mathrm{rec}$, independently of the initial number of plasmoids in the current sheet and the separation of scales. This is consistent with the number of hot spots we observe in the inner ring of the image.


\paragraph{Observability in the radio band}

The typical flux from the inner $10 \, \rg$ in M87* at the frequency $\nu_0 = \SI{230}{\giga\hertz}$ is $\SI{1}{\jansky}$~\citep{EHT_1}, translating into a radiated power $\nu F_\nu \approx 10^{40} \, \SI{}{\erg\per\second}$. The total jet power $L_\mathrm{jet}$ is likely to lie within the range $10^{42} - 10^{45} \, \SI{}{\erg\per\second}$~\citep{Broderick_2015a,Prieto_2016}. Assuming reconnection in the equatorial plane powers the emission, an upper bound on the total dissipated energy is $L_\mathrm{dis} \sim \beta_\mathrm{rec} L_\mathrm{jet}$, with $\beta_\mathrm{rec} \approx 0.1$ the collisionless reconnection rate~\citep{Sironi_2016}. The synchrotron cooling time scale $\tau_\mathrm{s}$ is related to the inverse Compton cooling time $\tau_\mathrm{IC}$ in the Thomson regime by $\tau_\mathrm{s} / \tau_\mathrm{IC} = U_\mathrm{IC} / \left( B_0^2 / 8 \pi \right) \approx 10^{-2}$, with $B_0 \approx \SI{10}{} - \SI{100}{\gauss}$ the typical magnetic field strength at horizon scale~\citep{EHT_2021b} and $U_\mathrm{IC} \approx 10^{-2} \, \SI{}{\erg\per\centi\meter\cubed}$ the energy density of low-energy photons in the inner zone ($\sim 10 \, \rg$)~\citep{Broderick_2015b,EHT_2021c}. Since $\tau_\mathrm{s} \ll \tau_\mathrm{IC} \ll \rg/c$, most of the magnetic energy dissipated by magnetic reconnection, initially carried by nonthermal particles, is converted into synchrotron emission.

The spectrum of reconnection-accelerated particles in pair plasma with zero guide field can largely be described as a power-law $\diff{N} / \diff{\gamma} \propto \gamma^{-p}$~\citep{Sironi_2014,Guo_2014,Werner_2016}. The index $p$ decreases with increasing upstream magnetization $\sigma$, going from $p \approx 2$ for $\sigma \gtrsim 10$ to $1$ at $\sigma \gg 10$, in the absence of synchrotron cooling~\citep{Werner_2016}. As $p$ gets closer to $1$, more and more energy is carried by high-energy particles. The synchrotron spectrum emitted by such a population of leptons also follows a power-law, with a spectral energy density depending on frequency as $\nu F_\nu \propto \nu^{(3-p)/2}$. The normalization is such that the total radiated power matches $L_\mathrm{dis}$. The peak of the total synchrotron spectrum cannot extend significantly beyond the burn-off limit at the photon energy $h \nu_\mathrm{rad} \approx \SI{20}{} \left( \beta_\mathrm{rec} / 0.1 \right) \SI{}{\mega\electronvolt}$, with $h$ the Planck constant ($\nu_\mathrm{rad} \sim 10^{10} \nu_0$). The characteristic synchrotron frequency of the photons emitted by the highest-energy particles $\nu_\mathrm{s} \sim \sigma^2 e B_0 / \me c$ lies above $\nu_\mathrm{rad}$ for $\sigma \gtrsim 10^6$. The value of $\sigma$ in M87* is likely to be much larger than $1$, but its exact value is uncertain.

If $\sigma \gtrsim \approx 10^6$, reconnection in the bare current sheet occurs in the strong cooling regime. Assuming $p\lesssim 2$, the radiated power $\nu F_{\nu}$ at the radio frequency $\nu_0 = \SI{230}{\giga\hertz}$ approximately reads $L_\mathrm{dis} (\nu_0 / \nu_\mathrm{rad} )^{(3-p)/2}$, yielding
\begin{equation} \label{eq:flux}
    \nu F_\nu \approx 10^{33} \left( \dfrac{L_\mathrm{jet}}{10^{44} \, \SI{}{\erg\per\second}} \right) \left( \dfrac{h \nu_\mathrm{rad}}{\SI{20}{\mega\electronvolt}} \dfrac{\SI{230}{\giga\hertz}}{\nu_0} \right)^{\textstyle \frac{p-1}{2}} \, \SI{}{\erg\per\second},
\end{equation}
where we have neglected gravitational redshift. Here, the most energetic photons have energies in the $\SI{}{\mega\electronvolt}$ range, so that they can produce pairs in the upstream~\citep{Hakobyan_2019,Mehlhaff_2021}. We show in the Supplemental Material that these secondary pairs also cool efficiently through synchrotron radiation, and emit in the radio band. Their total luminosity is $\SI{e39}{\erg\per\second}$, which is much larger than the estimate from Eq.~\eqref{eq:flux}, but is still outshone by the accretion disk, though marginally. Hence, the radio image would be dimmer in the ﬂaring state.


On the other hand, $\sigma$ could be smaller than $10^6$ but still large enough for reconnection to occur in the relativistic regime. Although the multiplicities resulting from electromagnetic cascades in the polar spark gaps are not very high~\citep{Chen_2020,Crinquand_2020}, additional mass loading from the accretion flow could reduce $\sigma$ with respect to the values defined by the plasma densities produced in these gaps. In that case, the particle spectrum retains its non-radiative shape approximately up to $\gamma \sim \mathrm{few} \times \sigma$. Beyond, the spectrum steepens, the exact shape depending on the relative strengths of cooling and secondary acceleration mechanisms. To give a simple estimate of the power at radio frequencies, we assume that most of the power is carried by particles below $\sigma$. The energetic constraint on the synchrotron radio flux is then relaxed with respect to the $\sigma > 10^6$ case: effectively, $\nu_\mathrm{rad}$ must be replaced by $\nu_\mathrm{s} \sim \sigma^2 e B_0 / \me c$ in Eq.~\eqref{eq:flux}. Fixing the power-law index at $p_0=1$, a typical value in the case of $e^\pm$ non-radiative reconnection up to $\sigma$, we obtain $\nu F_\nu \approx 10^{40} \left( L_\mathrm{jet} / 10^{44} \, \SI{}{\erg\per\second} \right) \left( \sigma / 10^3 \right)^{p_0-3} \, \SI{}{\erg\per\second}.$
The radiation emitted by the current sheet in the high-energy flaring state could then reach levels similar to the quiescent state and be observable by the EHT. To obtain this estimate, we have neglected the contribution of additional acceleration mechanisms, beyond the impulsive acceleration of particles by the reconnected electric field, to the energy budget~\citep{Petropoulou_2018, Hakobyan_2021, Zhang_2021}.


\paragraph{Conclusion}


In this letter, we have provided a physically motivated, first-principles model for the image of a low-luminosity AGN. The validity of our results is limited to describing a high-energy flaring state, in which the accretion flow has receded drastically, rather than the quiescent state observed by the EHT. However, even if the current sheet that we obtain from our GRPIC simulations results from idealized initial conditions, very similar configurations are reached by GRMHD simulations, which also include accretion physics. As long as the upstream magnetization is much larger than $1$, the morphology of the image should only depend on the black-hole spin.

Since Sgr A* is expected to vary on time scales of several minutes, future observations will be able to observe changes in the morphology of the image within the EHT observation window. From our simulations, we expect the radius of the bright ring to change with time, and possible hotspots to move along the ring, thereby providing a test of our model. In M87*, these hot spots should fully rotate along the ring in $5$ days, whereas in Sgr A* it would take about $5$ minutes. 

\begin{acknowledgements}

The authors would like to thank the anonymous referees for insightful comments, as well as Bart Ripperda and Fabio Bacchini for helpful discussions. This project has received funding from the European Research Council (ERC) under the European Union’s Horizon 2020 research and innovation programme (grant agreement No 863412). Computing resources were provided by TGCC and CINES under the allocation A0090407669 made by GENCI, and by Princeton Research Computing, a consortium of groups including the Princeton Institute for Computational Science and Engineering (PICSciE) and the Office of Information Technology's High Performance Computing Center and Visualization Laboratory at Princeton University.

\end{acknowledgements}

\widetext
\clearpage

\begin{center}
\textbf{\large Multi-dimensional simulations of ergospheric pair discharges around black holes:\\ 
Supplemental Material}
\end{center}

\section{Numerical techniques}

\subsection{Field equations and charge deposition}

In \texttt{GRZeltron}, we solve the electromagnetic field equations derived by \citet{Komissarov_2004} in the $3+1$ formalism:
\begin{eqnarray}
    \partial_t \vec{B} & = & - \Nabla \times \vec{E} \label{eq:maxwell_B},\\
    \partial_t \vec{D} & = & \Nabla \times \vec{H} - 4 \pi \vec{J}, \label{eq:maxwell_D}
\end{eqnarray}
where $\vec{H}$ and $\vec{E}$ are auxiliary fields defined by
\begin{eqnarray}
    \vec{H} & = & \alpha \vec{B} - \vec{\beta} \times \vec{D}, \\
    \vec{E} & = & \alpha \vec{D} + \vec{\beta} \times \vec{B}.
\end{eqnarray}
We have taken $c=1$ by convention. In these equations, $\alpha$ is the lapse function and $\vec{\beta}$ is the shift vector associated with a particular foliation of spacetime. We apply axial symmetry at the $\theta = 0$ and $\theta = \pi$ boundaries: $D^\varphi=0$, $B^\theta=0$ and $\partial D^r / \partial \theta = 0$. We use outflowing boundary conditions at the outer radial boundary, and zero-gradient conditions for $D^\theta$ and $B^r$ at the inner radial boundary, within the event horizon.

The current density $\vec{J}$ acts as a source term for the electromagnetic fields. It is obtained by depositing the current densities of macro-particles on the grid. We use a first-order interpolation of the particle coordinates. If a macroparticle with charge $q$ and weight $w$ is located between the nodes $r \in [r_i, r_{i+1}]$ and $\theta \in [\theta_j , \theta_{j+1} ]$, then the charge deposited on each neighboring node is
 \begin{alignat}{4}
  q_{i,j} & = \dfrac{V_{i+1,j+1}}{V} \, q \, w, & q_{i,j+1} & = \dfrac{V_{i+1,j}}{V} \, q \,w, \\
  q_{i+1,j} & = \dfrac{V_{i,j+1}}{V} \, q \, w, & q_{i+1,j+1} & = \dfrac{V_{i,j}}{V} \, q \, w,
 \end{alignat}
where we have defined $V = (r_{i+1} - r_i) (\theta_{j+1} - \theta_j)$, and $V_{i,j} = \abs{(r - r_i) (\theta - \theta_j)}$.

\subsection{Divergence cleaning}
 
This choice of deposition is conceptually simple and robust, but does not maintain the Maxwell-Gauss equation $\Nabla \cdot \vec{D} = 4 \pi \rho$ to machine precision (with $\rho$ the charge density deposited by the particle distribution). This lack of charge conservation could disrupt the simulation on long time scales. To alleviate this problem, we perform elliptic divergence cleaning every $N_c$ time steps, with $N_c = 25$ in our case. We solve Poisson's equation $\Delta  \Phi = -(4 \pi \rho - \vec{\nabla} \cdot \vec{D})$ for the correction $\Phi$ to the total electric potential (with $\vec{D}$ the electric field before the cleaning), and then correct the electric field by the amount $\delta \vec{D} = - \Nabla \Phi$. For example, a 2D discretization of the general-relativistic Maxwell-Gauss equation reads

\begin{align}
 \lp \Delta \phi \rp_{i,j} & = \dfrac{1}{\sh_{i,j}} \lp C_1 \lp \phi_{i+1,j} - \phi_{i,j} \rp + C_2 \lp \phi_{i-1,j} - \phi_{i,j} \rp + C_3 \lp \phi_{i,j+1} - \phi_{i,j} \rp + C_4 \lp \phi_{i,j-1} - \phi_{i,j} \rp \rp \\
 & = \dfrac{\lp D^r \sh \rp_{i+1/2,j} - \lp D^r \sh \rp_{i-1/2,j}}{{\Delta r} \sh_{i,j}} + \dfrac{\lp D^\theta \sh \rp_{i,j+1/2} - \lp D^\theta \sh \rp_{i,j-1/2}}{{\Delta \theta} \sh_{i,j}} - 4 \pi \rho_{i,j},
\end{align}
where the weight coefficients area are

\begin{align}
C_1 & = \dfrac{\sh_{i+1/2,j} h^{rr}_{i+1/2,j}}{{\Delta r}^2}, \quad C_2 = \dfrac{\sh_{i-1/2,j} h^{rr}_{i-1/2,j}}{{\Delta r}^2}, \\
C_3 & = \dfrac{\sh_{i,j+1/2} h^{\theta\theta}_{i,j+1/2}}{{\Delta \theta}^2}, \quad C_4 = \dfrac{\sh_{i,j-1/2} h^{\theta\theta}_{i,j-1/2}}{{\Delta \theta}^2}.
\end{align}

\section{Ray-tracing}

Let us consider a photon with $4$-momentum $p_\mu$. Geodesic motion in Kerr spacetime is completely integrable, since three conserved quantities can be attributed to each geodesic: the energy $E=-p_t$, the angular momentum $L=p_\varphi$, and the Carter constant $Q$, defined as~\citep{Carter_1968}

\begin{equation} \label{eq:carter}
 Q =  {p_\theta}^2 - \cos^2{\theta} \lp a^2 \dfrac{E^2 \rg^2}{c^2} - \dfrac{L^2}{\sin^2{\theta}} \rp.
\end{equation}
Null geodesics are independent of the photon energy $E$, so we define the dimensionless parameters $\ell= c L / E \rg$ and $q^2 = Q c^2 / E^2 \rg^2$. If a photon reaches spatial infinity, $E$ is its redshifted energy as measured by an observer at infinity. A photon trajectory is fully determined by a given $\lp \ell, q^2 \rp$, supplemented by the initial signs of $\dot{r}$ and $\dot{\theta}$. Two geodesics are associated with these parameters (determined by the choice of one of these two signs), whereas the other choice of sign determines the direction of travel along that geodesic. 

Let us consider a distant observer equipped with a screen, the center of which lies on the black hole. The screen is perpendicular to the line of sight of the distant observer to the black hole, which has an inclination $\alpha_\mathrm{obs}$ with respect to the spin axis. Let us assume that a geodesic with parameters $(\ell, q^2)$ reaches this screen at the position $(X,Y)$. The impact parameter $Y$ is the apparent displacement of the photon in the direction parallel to the spin axis of the black hole, whereas $X$ is the apparent displacement in the direction perpendicular to the projected axis. They are given by~\citep{Cunningham_1973}

\begin{align}
 \dfrac{X}{\rg} & = - \dfrac{\ell}{\sin{\alpha_\mathrm{obs}}}, \\
 \dfrac{Y}{\rg} & = \pm \sqrt{q^2 + a^2 \cos^2{\alpha_\mathrm{obs}} - \ell^2 \, \mathrm{cotan}^2{\alpha_\mathrm{obs}}}.
\end{align}
The sign of $Y$ depends on the initial sign of $\dot{\theta}$.

At every time step, each macro-particle can emit a macro-photon along its direction of motion, by virtue of relativistic beaming. To simplify matters, and because we do not claim to model realistic spectra given our limited scale separation, we assume that the power spectrum of any macro-photon is monoenergetic at the characteristic local synchrotron frequency $\nu_\mathrm{s}$. The total distribution of synchrotron photons is downsampled to maintain a reasonable computational cost. We assume that the plasma is optically thin everywhere and neglect annihilation between two synchrotron photons, so that once emitted they no longer interact with the plasma. As a result, their information can be saved and post-processed. Rather than the locally emitted frequency $\nu_\mathrm{s}$, we actually store the redshifted frequency $\nu_\infty$ as measured by an observer at infinity. It is computed as $h \nu_\infty = \alpha h \nu_\mathrm{s} - \beta^i p_i$, where $h$ is the Planck constant, $p_i$ the photon $3$-momentum, $\alpha$ the lapse function and $\beta^i$ the shift vector of the Kerr spacetime. 

The screen has a resolution $200 \times 200$ and a field of view $8 \, \rg \times 8 \, \rg$. The angular resolution is $9^\circ$ in the polar direction and $15^\circ$ in the azimuthal direction. To reduce shot noise, we apply a median filter on all images with a window of $3$ pixels. In all displayed images, we only select photons with frequencies between $e B_0 / \me c$ and $10 \, e B_0 / \me c$, a band that lies roughly in the middle of the simulated synchrotron spectrum. This allows us to focus on photons emitted by accelerated particles and to get rid of the diffuse background emission, while retaining a pass-band wide enough to have good statistics.

\section{Luminosity of secondary pairs if $\sigma > 10^6$}

We define $\gamma_\mathrm{rad}$ as the Lorentz factor of particles at which the synchrotron drag matches the accelerating electric force. $\gamma_\mathrm{rad}$ can therefore be expressed as $\beta_\mathrm{rec} B_0 = \sigma_\mathrm{T} U_\mathrm{B} \gamma_\mathrm{rad}^2$ (with $\sigma_\mathrm{T}$ the Thomson cross section and $U_\mathrm{B}$ the magnetic energy density), yielding $\gamma_\mathrm{rad} \approx 10^6$ in the case of M87*. In the regime where $\sigma > \gamma_\mathrm{rad}$, most of the dissipated energy is radiated close to the broad peak at the synchrotron burnoff limit at $\epsilon_\gamma \approx \SI{20}{\mega\electronvolt}$. Photons with such energy can annihilate with each other and produce secondary pairs with characteristic Lorentz factors $\gamma_\mathrm{s} \sim \epsilon_\gamma / \me c^2 \approx 100$. The synchrotron cooling time $t_\mathrm{syn}$ of these secondary pairs is given by

\begin{equation}
 \dfrac{c t_\mathrm{syn}}{\rg} \sim \dfrac{\me c^2}{\gamma_\mathrm{s} \sigma_\mathrm{T} \rg U_B},
\end{equation}
with $U_B$ the magnetic energy density. Even for secondary pairs, we have $c t_\mathrm{syn} / \rg \sim 10^{-3} \ll 1$, for a magnetic strength of $\SI{100}{\gauss}$. Consequently, secondary pairs radiate almost instantly their energy before they have time to escape. Most of this energy is emitted around $h \gamma_\mathrm{s}^2 e B_0 / \me c \approx \SI{10}{\milli\electronvolt}$, so that these pairs contribute to the $\SI{}{\milli\meter}$ wavelengths considered in this paper. 

The optical depth to pair production for photons at energy $\epsilon_\gamma$ can be estimated as 

\begin{equation}
 \tau_{\gamma\gamma} \sim \sigma_{\gamma\gamma} l \dfrac{U_\mathrm{s}\left(\epsilon_\mathrm{s} \right)}{\epsilon_\mathrm{s}},
\end{equation}
where $l \sim 10 \, \rg$ is the typical size where high-energy photons are produced, $\sigma_{\gamma\gamma} \sim 0.2 \, \sigma_\mathrm{T}$ is the peak cross section for photon annihilation, and $U_\mathrm{IC} \left(\epsilon_\mathrm{s} \right)$ is the energy density of photons at the target energy $\epsilon_\mathrm{s} = \left( \me c^2 \right) / \epsilon_\gamma$ (with $\epsilon_\mathrm{s} \approx \epsilon_\gamma$ in the case of photon-photon annihilation in the $\SI{}{\mega\electronvolt}$ range). Estimating $U_\mathrm{s}\left(\epsilon_\mathrm{s} \right) = L_\mathrm{dis} / 4 \pi l^2 c$, we obtain $\tau_{\gamma\gamma} \sim 10^{-4}-10^{-5} \left( L_\mathrm{jet} / 10^{44} \, \SI{}{\erg\per\second} \right) \left( 10 \, \rg / l \right)$~\citep{Ripperda_2022}. In this optically thin regime, the pair creation rate is then given by $\dot{N} \simeq \tau_{\gamma\gamma} L_\mathrm{dis} / \epsilon_\gamma$. Because the energy in the secondary pairs is quickly radiated away, the secondary pairs synchrotron luminosity is given by $L_\mathrm{sec} \sim \gamma_\mathrm{s} \me c^2 \dot{N} \sim \tau_{\gamma\gamma} L_\mathrm{dis}$. For $L_\mathrm{jet} = 10^{44} \, \SI{}{\erg\per\second}$, we obtain $L_\mathrm{sec} \approx 10^{39}-10^{40} \, \SI{}{\erg\per\second}$.

\bibliographystyle{apsrev4-2}

\bibliography{biblio}

\end{document}